\documentclass[reprint,amsmath,amssymb,aps,prl]{revtex4-1}
\usepackage{graphicx}% Include figure files
\usepackage{dcolumn}% Align table columns on decimal point
\usepackage{bm}% bold math
\usepackage{textcomp}

\usepackage{color,soul}

\pdfoutput=1

\begin{document}

\preprint{APS/123-QED}

\title{Low-energy electron escape from liquid interfaces: charge and quantum effects}

\author{Loren Ban}
\author{Thomas E. Gartmann}%
\author{Bruce L. Yoder}%
\author{R. Signorell}%
\email[email: ]{rsignorell@ethz.ch}%Lines break automatically or can be forced with \\
\affiliation{%
	Department of Chemistry and Applied Biosciences, ETH Z{\"u}rich, \\Vladimir-Prelog Weg 2, CH-8093 Z{\"u}rich, Switzerland
}%

\date{\today}% It is always \today, today,
             %  but any date may be explicitly specified

\begin{abstract}
The high surface sensitivity and controlled surface charge state of submicron sized droplets is exploited to study low-energy electron transport through liquid interfaces using photoelectron imaging. Already a few charges on a droplet are found to modify the photoelectron images significantly. For narrow escape barriers, the comparison with an electron scattering model reveals pronounced quantum effects in the form of above-barrier reflections at electron kinetic energies below about 1 eV. The observed susceptibility to the characteristics of the electron escape barrier might provide access to these properties for liquid interfaces, which are generally difficult to investigate.
\end{abstract}

%\keywords{Suggested keywords}%Use showkeys class option if keyword
                              %display desired
\maketitle

%\tableofcontents

The interaction of low-energy electrons (LEE, $< 50$ eV) with condensed molecular matter is relevant to many fields, ranging from radiation damage of biological systems, to atmospheric chemistry and astrochemistry, to the engineering of electronic devices \cite{Alizadeh2015,Michaud2003,Lu2001,Boyer2016,Ferradini1991,Naaman2007}. A phenomenon that has received particularly broad attention in this context is the formation of the solvated electron and the role it might play in radiation damage \cite{Young2012,Elkins2013a,Luckhaus2017g,Karashima2019,Ma2009,Lietard2019,Gartmann2019,Turi2012,Herbert2017,Zho2018}. Experimental studies mainly concentrated on low-energy photoelectron transmission (LEPET) or low-energy electron transmission (LEET) spectroscopy of thin films \cite{Naaman2007}, while liquid microjets, aerosol particles, and molecular clusters were later suggested as alternative samples for the investigation of LEE transport in dielectrics \cite{Luckhaus2017g,Ziemann1995,Ziemann1996,Signorell2016,Wilson2007,Goldmann2015,Amanatidis2017,Jacobs2017,Seiiert2017,Thurmer2013a,Suzuki2014,Nishitani2017,Zhang2013a,Gartmann2018,Hartweg2017b}. 

Interfacial electron transfer and the scattering processes it involves play a vital role in LEE transport \cite{Ferradini1991,And1998,Hiraoka1981,Nitzan2001,Herring1949,Cutler1964,Crowell1966,Bouchard1980,Jay-Gerin1983,Harris1997}, but they are particularly difficult to investigate experimentally in the case of (volatile) molecular liquids. Major obstacles arise from radiation-induced charging and the incompatibility of high-vacuum conditions with thin film and bulk samples of high vapor pressure. As a result, there is still no consensus even on the liquid-vacuum interface potential of water \cite{Goulet1990,Coe1997,Chen2016a,Gaiduk2018,Elles2006,Bernas1997}, with reported values of the escape barrier typically varying between 0.1 and 1.2 eV. The situation is further complicated by the sensitivity of LEE escape from interfaces to the presence of even only a few charges. Corresponding studies are rather scarce \cite{Marsolais1989,Michaud2007}, presumably because it is difficult to control the exact charge state of thin-film or bulk samples. Small particles have been used to study electron impact or photoelectron charging mechanisms \cite{Ziemann1995,Ziemann1996,Burtscher1982a,Nishida2017,Grimm2006}, but to the best of our knowledge the LEE escape from the interface has not been investigated in detail.

Here, we present the results of a combined experimental and theoretical study of LEE transfer across liquid-vacuum interfaces. We use angle-resolved photoelectron spectroscopy of submicron-sized droplets to access information on the photoelectron kinetic energy (eKE) and the photoelectron angular distribution (PAD). Droplets offer important advantages over thin film or bulk samples for such studies: (i) vacuum compatibility even for volatile molecular liquids, (ii) no radiation-induced sample charging because of constant sample refreshment, (iii) high surface sensitivity, and (iv) control over and independent determination of the charge state. This allows us to exploit the high sensitivity of LEE below a few eV to the properties of the interface potential and charge state. The comparison with an electron scattering model reveals that the photoelectron distribution is not only affected by the electrostatic interaction with the droplet charge. Depending on the characteristics of the interface potential quantum effects can play a significant role in LEE transfer through the liquid-vacuum interface.

\section{Experiment}
 Liquid dioctyl phthalate (DEHP) droplets were produced by atomization, size-selected in a differential mobility analyzer and charged in a unipolar charger. Droplets with an average radius $<R_D>$ of $\sim$210 nm were investigated in five different average charge states $<q>=$ +16, +8, 0, -7, -15  (Figs. S3-S6 \cite{SIref}). The droplets were then transferred to vacuum by an aerodynamic lens (ADL) \cite{Liu1995,Liu1995a} and resonantly 2-photon ionized by a 266 nm nanosecond laser. Photoelectron eKE and angular distributions were recorded with a velocity map imaging (VMI) spectrometer \cite{Yoder2013}. PADs were analyzed directly in terms of the raw images, while eKE spectra were retrieved from photoelectron images reconstructed with MEVIR \cite{Dick2014} (see Ref. \cite{SIref}).

\section{Electron scattering model}
\begin{figure}[h]
	\includegraphics[width=0.9\columnwidth]{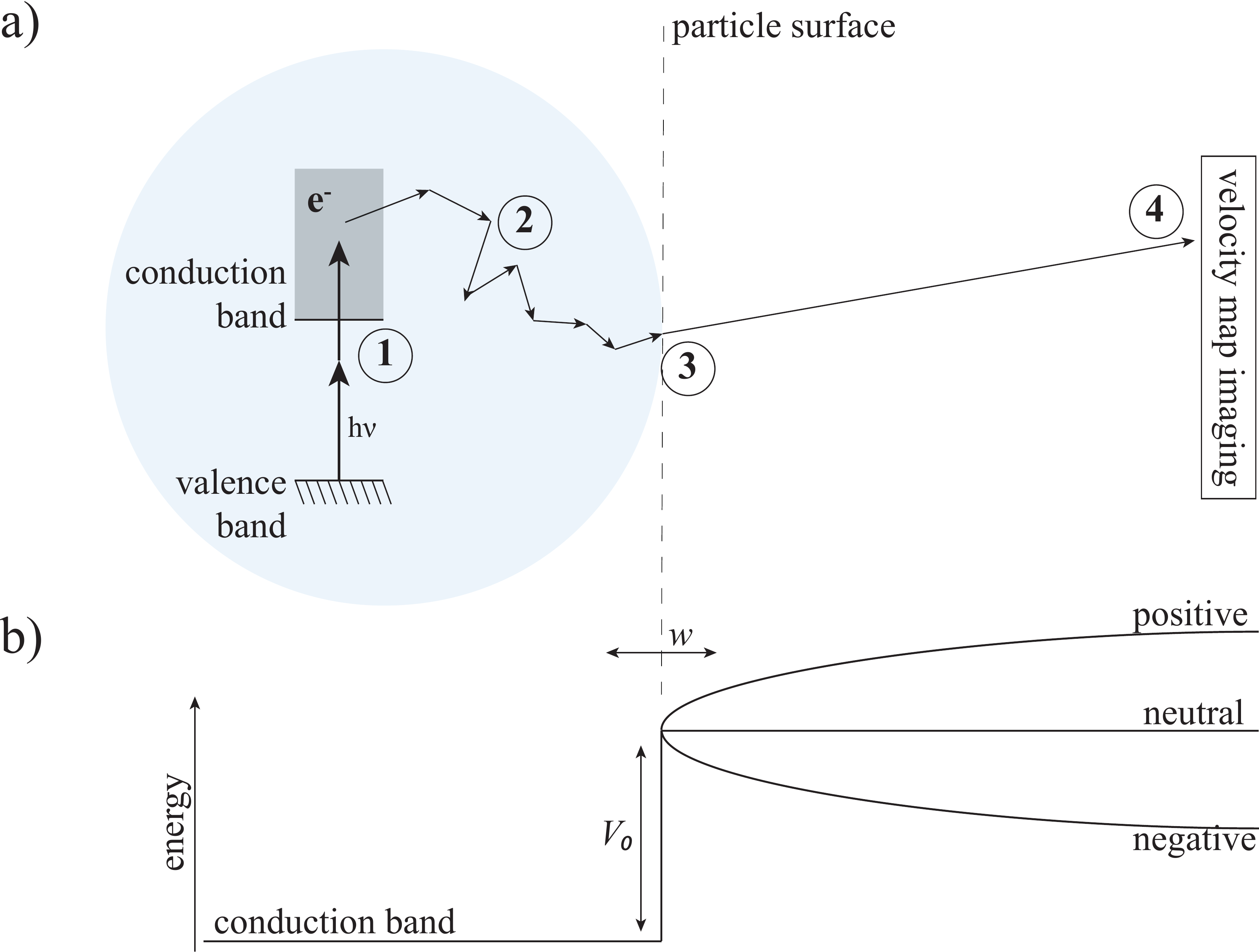}% Here is how to import EPS art
	\caption{\label{fig:potential} (a) Photoionization (1), electron transport scattering (2), escape at the droplet-vacuum interface (3), and VMI detection (4). (b) Charge dependent potential energy (excluding centrifugal potential \cite{SIref}).}
\end{figure}

 The model for the photoionization (step 1 in Fig. \ref{fig:potential}a), electron transport scattering (step 2), electron escape at the droplet-vacuum interface (step 3), and detection by VMI (step 4) follows our previous work \cite{Goldmann2015,Signorell2016,Hartweg2017b,Luckhaus2017g} with extensions for the treatment of surface charges and electron escape at the droplet-vacuum interface. A laser excites valence electrons into the conduction band (step 1). The genuine binding energy spectrum of DEHP required to describe this step was determined from a fit to the experimental spectrum of the neutral particles ($<q>\,=0$ in Fig. \ref{fig:results}b) excluding eKEs$\,\leq 0.1$ eV \cite{SIref}. The obtained genuine spectrum (Fig. S7 \cite{SIref}) is in reasonable agreement with the calculated gas-phase spectrum of dimethyl phthalate with a gas-to-liquid shift of $\sim2$ eV, similar to benzene \cite{Schwarz1982,Scott1982,Saik1994}. Electron transport scattering (step 2) was modeled with a probabilistic electron scattering model \cite{Signorell2016,Hartweg2017b,Luckhaus2017g}, which amounts to a Monte Carlo solution of the transport equation. The requisite differential scattering cross sections (Fig. S8 \cite{SIref}) were derived from data for benzene \cite{Goulet1985,Goulet1986} as described in Ref. \cite{SIref}. 
 We assume charges to be uniformly distributed on the droplet's surface adding a constant potential inside the droplet \cite{Ziemann1995} and a Coulomb potential outside to the neutral droplet's step-like barrier at the interface with height $V_0$ and width $w$ (Fig. \ref{fig:potential}b).

 To account for quantum effects we calculate the transmission probability $T$ at the interface from the numerical solution of the radial Schr{\"o}dinger equation for the given potential. As a consequence of centrifugal potential contributions, $T$ depends on the impact angle $\theta$ of the electron relative to the surface normal (Fig. \ref{fig:reflection}c and Ref. \cite{SIref}).  Following transmission, electrons are propagated classically before their final velocity is projected onto the detector plane to produce the image (VMI). Images simulated for different sizes and charge states are averaged according to the experimentally determined size and charge distributions \cite{SIref}.

\section{Results and discussion}
\begin{figure}[h]
	\includegraphics{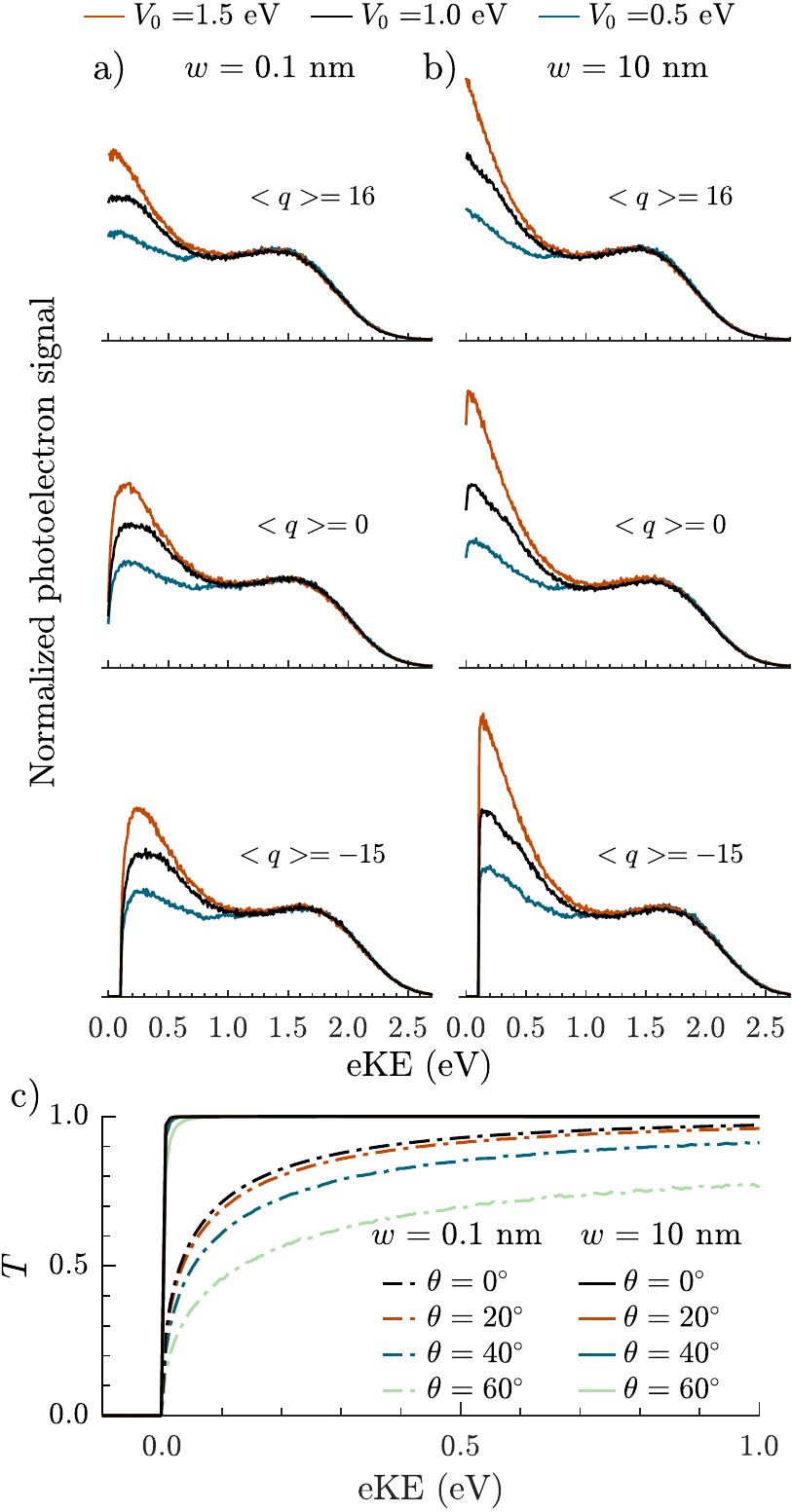}% Here is how to import EPS art
	\caption{\label{fig:reflection} Simulated eKE spectra for (a) $w=0.1$ nm and (b) $w=10$ nm. Blue line: $V_0=$ 0.5 eV, Black line: $V_0=$ 1.0 eV, Red line: $V_0=$ 1.5 eV. The same genuine properties are used for all simulations. (c) Electron transmission coefficient $T$ for $V_0=$ 1.0 eV and $<q>=0$ as a function of the eKE for $w=0.1$ nm (dotted lines) and $w=10$ nm (solid lines) for different incidence angles $\theta$ (colors).}
\end{figure}

 Fig. \ref{fig:results}a shows experimental photoelectron VMIs (top row) for droplets with average charge states $<q>=$~+16 (left), 0 (center) and -15 (right), respectively. As a result of nanofocusing, the images show strong asymmetry along the laser propagation direction with higher electron intensity opposite the illuminated side of the droplet \cite{Signorell2016}. Variation of the charge state from positive to negative decreases  the near zero eKE electron signal (near the image center) and increases that of  higher eKE electrons (towards the image border). This is accompanied by a change in the angular distribution of the electron signal. The trends become clearer in the eKE spectra (Fig. \ref{fig:results}b, top row) and PADs (Fig. \ref{fig:results}c, top row) retrieved from the images. Generally, good agreement is found between experiments and scattering simulations (Figs. \ref{fig:results}a,b,c, bottom row). Based on their charge-dependence, the spectra can be divided into a high ($>1$ eV) and a low energy ($<1$ eV) region:
\begin{figure}
	\includegraphics[width=0.9\columnwidth]{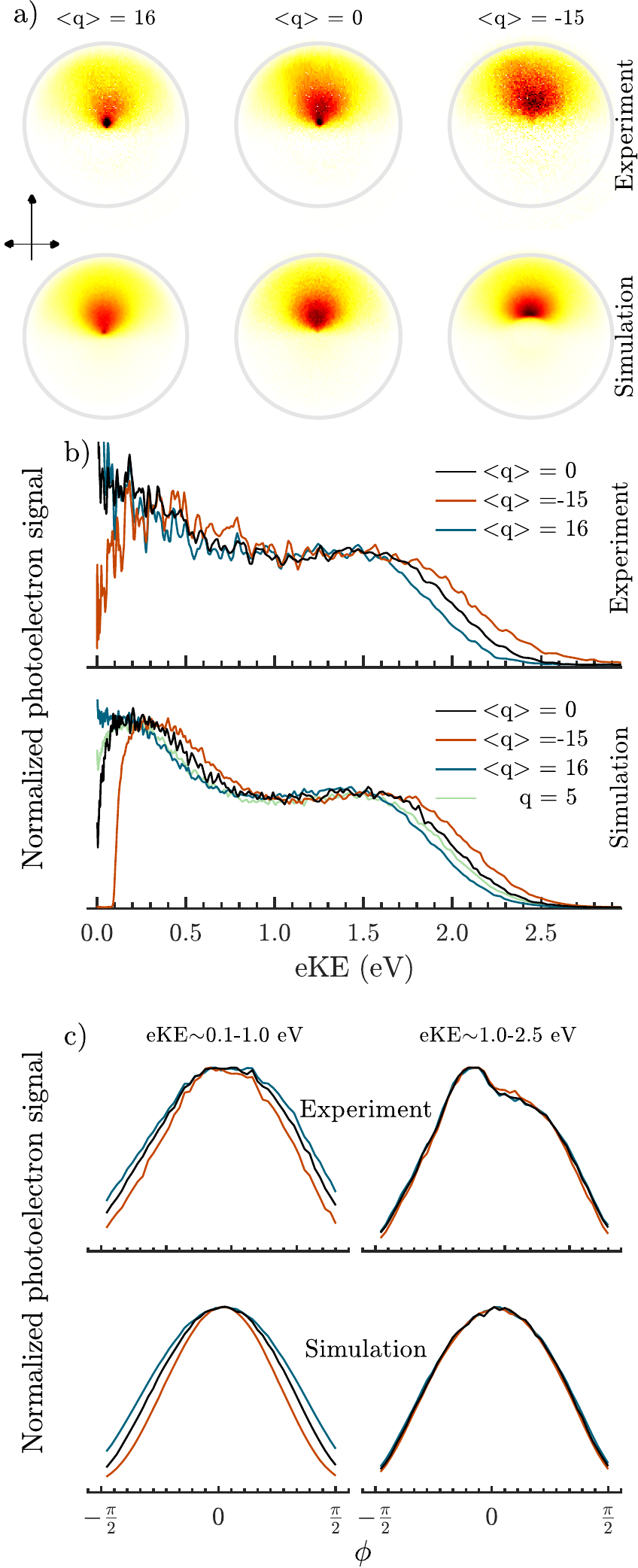}% Here is how to import EPS art
	\caption{\label{fig:results} (a) Experimental (top) and simulated (bottom) velocity map images of DEHP droplets. The arrows indicate the laser propagation and polarization direction. The simulations are for $V_0=1$ eV and $w=0.1$ nm and a genuine photoelectron spectrum obtained by a fit to the neutral experimental eKE spectrum. (b) eKE spectra retrieved for the neutral (black), positively (blue) and negatively charged droplets (red). Additional simulation for a charge of $q=5$ (green). The spectra are normalized to their respective values at 1.5 eV. (c) Angular distribution of the integrated low (left) and high (right) eKE signal (see text and  \cite{SIref}). $\phi$ is the angle w.r.t. the direction of light propagation. Distributions are normalized to their respective values at $\phi=0$.}
\end{figure}
\paragraph*{High \textnormal{eKE} region ($>1$ \textnormal{eV}):} The main feature in the eKE spectra (Fig. \ref{fig:results}b) is the charge-dependent energy shift of the high eKE onset (Table \ref{tab:charge_shift} and Ref. \cite{SIref}), which reflects the corresponding shift of the vacuum level (Fig. \ref{fig:potential}). Relative to the neutral case, spectra of positively charged droplets are shifted towards lower eKE by up to 100 meV, while we observe shifts to higher eKE of up to 250 meV for negative charges. Most of the observed shifts of the high eKE edge agree with the corresponding calculated average shift of the vacuum level which is proportional to $<q>$ (Fig. S9 \cite{SIref}). In this region the electron's kinetic energy significantly exceeds both the barrier height and even more so the Coulomb shift of the vacuum level. The higher the eKE value lies above the top of the barrier, the more classical the transmission becomes. As a consequence the shape of the high eKE edge is neither pronouncedly influenced by quantum effects (see below) nor by the shape of the barrier ($V_0$ and $w$). The latter is illustrated by the simulations for different values of $V_0$ and $w$ which leave the high eKE region unaffected (Fig. \ref{fig:reflection}). We note that changing $V_0$ or $w$ cannot shift the photoelectron spectrum on the eKE axis (as the surface charge does) but only affect relative intensities in the spectrum.

Good agreement between experimental and simulated PADs is reflected in the velocity map images (Fig. \ref{fig:results}a). This is illustrated in Fig. \ref{fig:results}c for the charge dependent angular distribution of the integrated low (left column) and high (right column) eKE signal. The traces were obtained by integrating the images for each given direction over electron velocities in the detector plane corresponding to eKEs in the range 0.1-1.0 eV and 1.0-2.5 eV, respectively. The direction is specified in terms of the angle $\phi$ w.r.t. the direction of propagation. Contrary to the eKE spectra, no significant charge dependence is observed in the high eKE range between 1.0-2.5 eV (Fig. \ref{fig:results}c, right column and Fig.S10 in\cite{SIref}). (The  asymmetry w.r.t. $\phi=0$ in the experimental images for eKEs between 1.0-2.5 eV is due to inhomogeneities of the electron imaging detector.)  
\begin{table}[h]
	\caption{\label{tab:charge_shift}%
		Charge-dependent high kinetic energy onset (eKE onset) determined at the 1/$e^2$ signal level \cite{SIref} and energy shift $\Delta$eKE relative to the neutral case.}
	\begin{ruledtabular}
		\begin{tabular}{ccccc}
			$<q>$ & \multicolumn{2}{c}{eKE onset (eV)}& \multicolumn{2}{c}{$\Delta$eKE (eV)}\\		
			& experiment & simulation & experiment & simulation\\
			\hline
			16 & 2.21 & 2.21 & -0.11 & -0.10 \\ 			
			8 & 2.26 & 2.25 & -0.06 & -0.06\\ 
			0 & 2.32 & 2.31 & 0 & 0 \\
			-7 & 2.40 & 2.37 & 0.08 & 0.06\\
			-15 & 2.57 & 2.43 & 0.25 & 0.12 \\
		\end{tabular}
	\end{ruledtabular}
\end{table}

\paragraph*{Low \textnormal{eKE} region ($<1$ \textnormal{eV}):} Negative charges lead to significantly lower electron signals at low eKE compared with neutral and positive charges (Fig. \ref{fig:results}b). Even though this general trend is captured by the simulation, there are deviations between simulation and experiment, in particular for the neutral droplets. They can be partly attributed to the larger experimental uncertainties in this low eKE range and to difficulties in generating neutral droplets. Compared with higher eKE electrons, it is more challenging to quantitatively record very low eKE electrons, which are much more sensitive to small perturbations (e.g. external fields, imperfections in the VMI optics) affecting the measured eKE values. Furthermore, the lower angular resolution in the center of the electron detector strongly reduces the signal to noise level for near zero eKE electrons. Additional limitations arise for the neutral case, where the simulation predicts a smaller near zero eKE signal than found in the experiment. It is challenging to generate completely uncharged droplets, and difficult to quantify the exact charge state of droplets with very few charges on them. All this is exacerbated by the high sensitivity of the photoelectron spectrum in the low eKE region to the presence of even a small number of charges. This is illustrated by the simulation for $q=5$ in Fig. \ref{fig:results}b (green line). Already a small amount of positive charges considerably increases the signal near zero eKE compared with uncharged particles. 

The observed charge-dependence of the spectra in the low eKE region results from the electrostatic interactions of the electrons with the charged droplet surface following the escape. The main effect is the overall shift of the photoelectron spectrum already discussed above for the high kinetic energy edge. In contrast to high eKE electrons, the relatively large deBroglie wavelength ($\sim$1.7 nm at 0.5 eV) of low eKE electrons makes them very sensitive to quantum effects and the exact shape ($V_0$ and $w$) of the interface potential. Simulated eKE spectra for different barrier heights ($V_0=$ 0.5, 1.0, and 1.5 eV)  are shown in Fig. \ref{fig:reflection}a and b for two limiting cases of the barrier widths $w=$ 0.1 and 10 nm, respectively. The latter is certainly broader than would appear physically plausible. Here it only serves to illustrate a limiting (classical) behavior. A higher barrier is equivalent to a lower bottom of the conduction band and hence higher eKE of the electron inside the droplet. In this way, the relative abundance of low eKE electrons increases with $V_0$ as a consequence of the energy-dependence of the electron scattering cross sections (Fig. S7 \cite{SIref}). A narrow, step-like barrier ($w=0.1$ nm, Fig. \ref{fig:reflection}a) represents a case where quantum effects on the electron escape play a prominent role. These effects are suppressed in the case of a broad, smooth barrier ($w=10$ nm, Fig. \ref{fig:reflection}b). Irrespective of the barrier height $V_0$, the relative abundance of low eKE electrons is lower for the narrow barrier ($w=$ 0.1 nm), where quantum (above-barrier) reflections reduce the electron transmission through the interface. For a fixed barrier height $V_0=$ 1 eV and charge state $<q>=0$, Fig. \ref{fig:reflection}c shows the transmission coefficient $T$ as a function of eKE for $w=$ 0.1 nm (dashed lines) and $w=$ 10 nm (full lines) for different incidence angles $\theta$ (angle relative to the surface normal). Quantum reflections in the case of the narrow barrier strongly reduce  $T$ for eKE $<$ 1eV, while the broad barrier produces an almost classical transmission behavior. This holds regardless of the droplet charge, though it is most evident for the negatively charged droplets, where the sharp signal cutoff for $w=10$ nm transforms into a smoother decrease for $w=0.1$ nm (Fig. \ref{fig:reflection}a and b). With increasing barrier height $V_0$, the difference between the eKE spectra for $w=0.1$ and 10 nm becomes more pronounced as a result of the stronger reduction of $T$ for higher barriers (Fig. S11 and S12 \cite{SIref}). Significant contributions to the electron transmission arising from tunnelling (as opposed to quantum reflections) are not found in the simulations, as expected for the relatively broad potential cusp at the barrier (Fig. \ref{fig:potential}). 

In contrast to the high eKE region ($>1$ eV, Fig. \ref{fig:results}c, right), the angular distributions of low eKE signal ($<1$ eV, Fig. \ref{fig:results}c, left) is sensitive to the droplet charge. Fig. S11 \cite{SIref} reveals that the highest sensitivity arises for the lowest eKE values ($<0.5$ eV); i.e. the region where quantum transmission effects are pronounced (Fig. \ref{fig:reflection}c). The general trend is a reduction of the degree of forward scattering from negative to neutral to positive charge. Most of the effect has a classical origin: electrons are accelerated (or decelerated) by the Coulomb potential. If they exit the droplet at a non-vanishing angle to the surface normal (impact angle $\theta$, finite angular momentum) negative charges (acceleration) will reduce and positive charges (deceleration) increase that angle compared with the neutral case. As most electrons are ejected in the direction of propagation, negative (positive) charges will on balance reduce (increase) the relative abundance of electrons moving orthogonal to the direction of propagation, which is what we observe. Although largely classical this effect is enhanced by above-barrier reflection, which preferentially reduces the transmission of electrons with large impact angles $\theta$. This quantum effect becomes more pronounced in going from positive to negative charging of the droplets ( $\theta=0$ in Fig. S11 vs. $\theta=60^\circ$ in Fig. S12 \cite{SIref}).

\section{Conclusion}
Angle-resolved photoelectron spectroscopy of neutral, positively and negatively charged droplets at electron kinetic energies below a few eV provides information on the electron escape from liquid interfaces. From the position of the high kinetic energy onset of the electron signal, we can determine charge-dependent binding (ionization) energies - which have been elusive to experimental probes for liquids. At very low kinetic energies ($<1$ eV) the photoelectron spectra  are highly sensitive to the exact charge state as well as the height and width of the electron escape barrier. The quantum effect of above-barrier reflections at narrow barriers strongly reduces the abundance of very low kinetic energy electrons compared with broader barriers. The comparison with an electron scattering model shows that tunneling - in contrast to reflection - does not significantly contribute to the observed photoelectron spectra. So far, consistent experimental values of barrier heights and widths are not available for most liquids, including water. The droplet approach might provide new experimental access to escape barrier properties even for such liquid interfaces. Furthermore, droplets offer a way to systematically investigate the influence of the charge state on the photoelectron spectra of liquids - an issue that is being discussed with regard to liquid-microjet photoelectron spectroscopy and more generally in the context of radiation-induced sample charging in condensed phase photoelectron spectroscopy.  
\\

\begin{acknowledgments}
We are very grateful to Dr. Martin Fierz for lending us a unipolar aerosol charger and for his advice concerning the aerosol charging experiments, and to Dr. David Luckhaus for his advice regarding the simulations. We thank David Stapfer, Markus Steger and Daniel Zindel for technical support. This project has received funding from the European Union’s Horizon 2020 research and innovation program from the European Research Council under the Grant Agreement No 786636, and from the Swiss National Science Foundation (SNSF) through SNSF project no. $200020\_172472$.
\end{acknowledgments}

\bibliographystyle{apsrev4-1} % Tell bibtex which bibliography style to use
\bibliography{bib_1}% Produces the bibliography via BibTeX.

% The \nocite command causes all entries in a bibliography to be printed out
% whether or not they are actually referenced in the text. This is appropriate
% for the sample file to show the different styles of references, but authors
% most likely will not want to use it.
%\nocite{*}

\end{document}

\preprint{APS/123-QED}
	
	\title{Supporting Information: Low-energy electron escape from liquid interfaces: charge and quantum effects}
	
	\author{Loren Ban}
	\author{Thomas E. Gartmann}%
	\author{Bruce L. Yoder}%
	\author{R. Signorell}%
	\email[email: ]{rsignorell@ethz.ch}%Lines break automatically or can be forced with \\
	\affiliation{%
		Department of Chemistry and Applied Biosciences, ETH Z{\"u}rich, \\Vladimir-Prelog Weg 2, CH-8093 Z{\"u}rich, Switzerland
	}%
	
	\date{\today}% It is always \today, today,
	%  but any date may be explicitly specified

\maketitle

\renewcommand{\thefigure}{S\arabic{figure}}% add prefix to figure captions
\onecolumngrid
\section{Experiment \label{sec:SIexp}}
\subsection{Experimental setup}
\begin{figure}[h]
	\includegraphics[width=0.9\columnwidth]{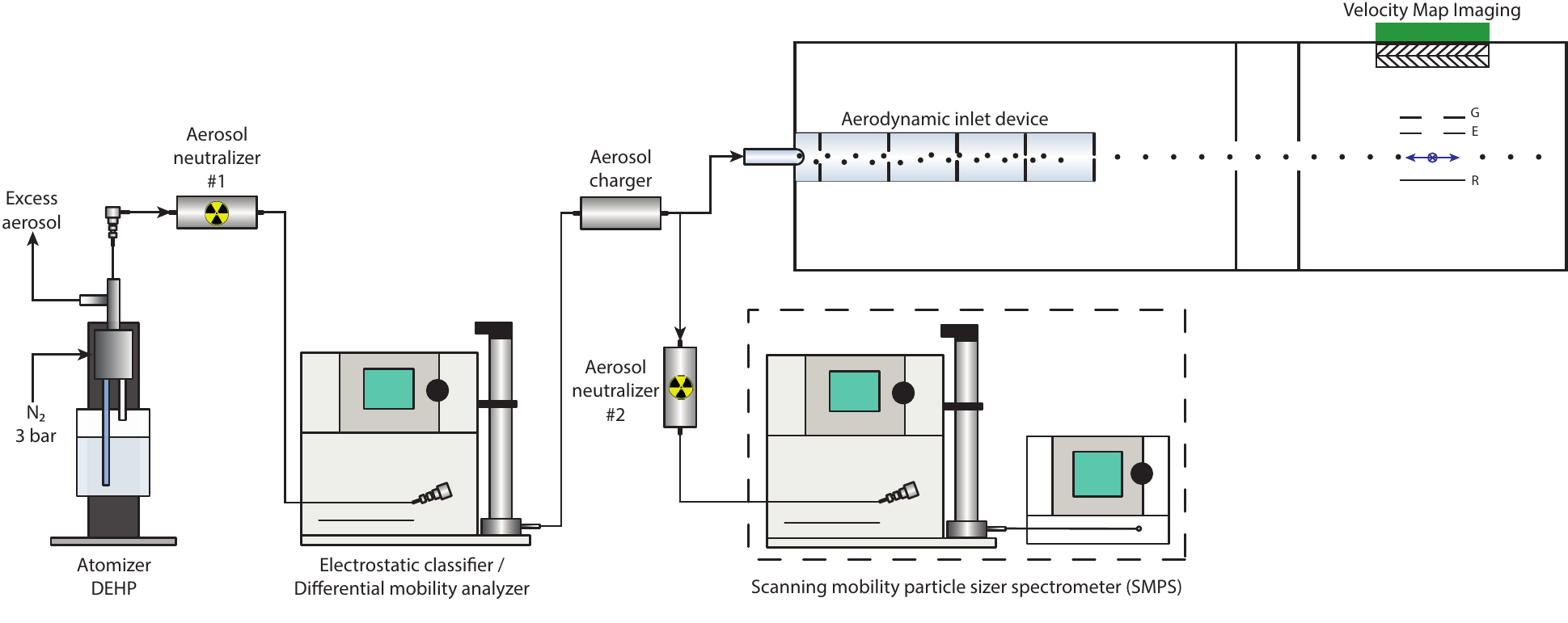}% Here is how to import pdf art
	\caption{\label{fig:experiment} Experimental setup used for measuring charge-dependent photoelectron spectra of submicron-sized DEHP droplets. R, E and G stand for the repeller, extractor and ground plates of the VMI optics, respectively.}
\end{figure}

The experimental setup used in this work is shown in Fig. \ref{fig:experiment}. It is comprised of two main parts; the air-side aerosol setup and the photoelectron spectrometer setup.

\paragraph*{Air-side aerosol setup:}
Bis(2-ethylhexyl) phthalate (DEHP, often named simply dioctyl phthalate) droplets are generated in a collision-type atomizer (TSI Model 3076) using 3 bar of N$_2$. The generated aerosol flows through an impactor (d=0.071 cm) at $\sim$1 L/min resulting in a cut-off diameter of $\sim800$ nm. The aerosol is then charge-equilibrated in an Xray neutralizer (TSI Model 3088) and sent through a differential mobility analyzer (Long DMA, TSI Model 3081) for size selection. A home-built corona-wire unipolar aerosol charger is used to charge the droplets in a controlled way, without altering the droplet size and chemical composition (confirmed by gas chroatography-mass spectrometry (GC-MS) analysis.). A scanning mobility particle sizer spectrometer (SMPS, Models 3936 and 3938) is used to obtain size and charge information in parallel to the photoelectron images.

\paragraph*{Photoelectron spectrometer:}
The droplets are transferred into vacuum with an aerodynamic lens (ADL) \cite{Signorell2016,Amanatidis2017}. The design (consisting of 5 orifices guiding and focusing the aerosol flow into the ionization region) is based on the work by McMurry and coworkers \cite{Liu1995,Liu1995a}. Droplets are resonantly 2-photon ionized at a wavelangth of 266 nm with a pulsed ns-laser operating at 20 Hz repetition rate (Quantel Ultra). The generated photoelectrons are recorded in a VMI photoelectron spectrometer \cite{Yoder2013,Signorell2014a,Goldmann2015,Signorell2016,Amanatidis2017} resulting in 2-dimensional photoelectron images. The electron kinetic energy (eKE) spectra are retrieved after reconstructing the images with MEVIR \cite{Dick2014}. The exact reconstruction requires cylindrical symmetry of the true 3-dimensional photoelectron distribution about the axis of reconstruction. In the droplets this is approximately fulfilled around the axis of light propagation, although the symmetry is slightly broken by the linear polarization of the ionizing laser radiation. However, this effect is marginal as any genuine anisotropy of the electron w.r.t the polarization of the laser tends to be averaged out by scattering. It has previously been shown for small droplets \cite{Amanatidis2017} that reconstructing along the laser propagation direction yields eKE spectra very close to the true eKE spectra. In the present work we have confirmed this by simulations, which show that the eKE spectrum obtained from the simulated 3-dimensional eKE distribution is essentially indistinguishable from the eKE spectrum retrieved from the simulated VMI by reconstruction along the propagation axis. 
The photoelectron angular distributions (PADs) shown in Fig. 2c in the main text were obtained by radially integrating the velocity-map images, as explained in Section \ref{sec:SIangular}.

\clearpage
\subsection{Determination of droplet size and droplet charge \label{sec:SIsize_charge}}

The main highlight of the setup in Fig. \ref{fig:experiment} is that it allows for in-parallel determination of the droplet charge- and size-distributions while performing photoelectron spectroscopy. In this way, direct assignment of each photoelectron image to a droplet sample of known size and charge is possible. The size and charge determination can be performed interchangeably by turning the aerosol neutralizer \#2 on and off (Fig. \ref{fig:experiment}), respectively. In the present study, the droplet charge is varied in a controlled way by adjusting the charger setting without disturbing the aerosol flow. One has to point out, however, limitations of the setup due to inherent incompatibilities of the high-vacuum photoelectron spectrometer with air-side aerosol equipment. Namely, high initial droplet concentration is required to obtain good signal from the size-selected and charged droplets in the spectrometer, leading to non-optimal conditions in the DMA and the Xray neutralizer \#1. These are manifest, for example, in a fairly broad droplet size distribution after the DMA. Further details on the droplet size and charge determination are given below. 

\begin{figure}[b]
	\includegraphics[width=0.6\columnwidth]{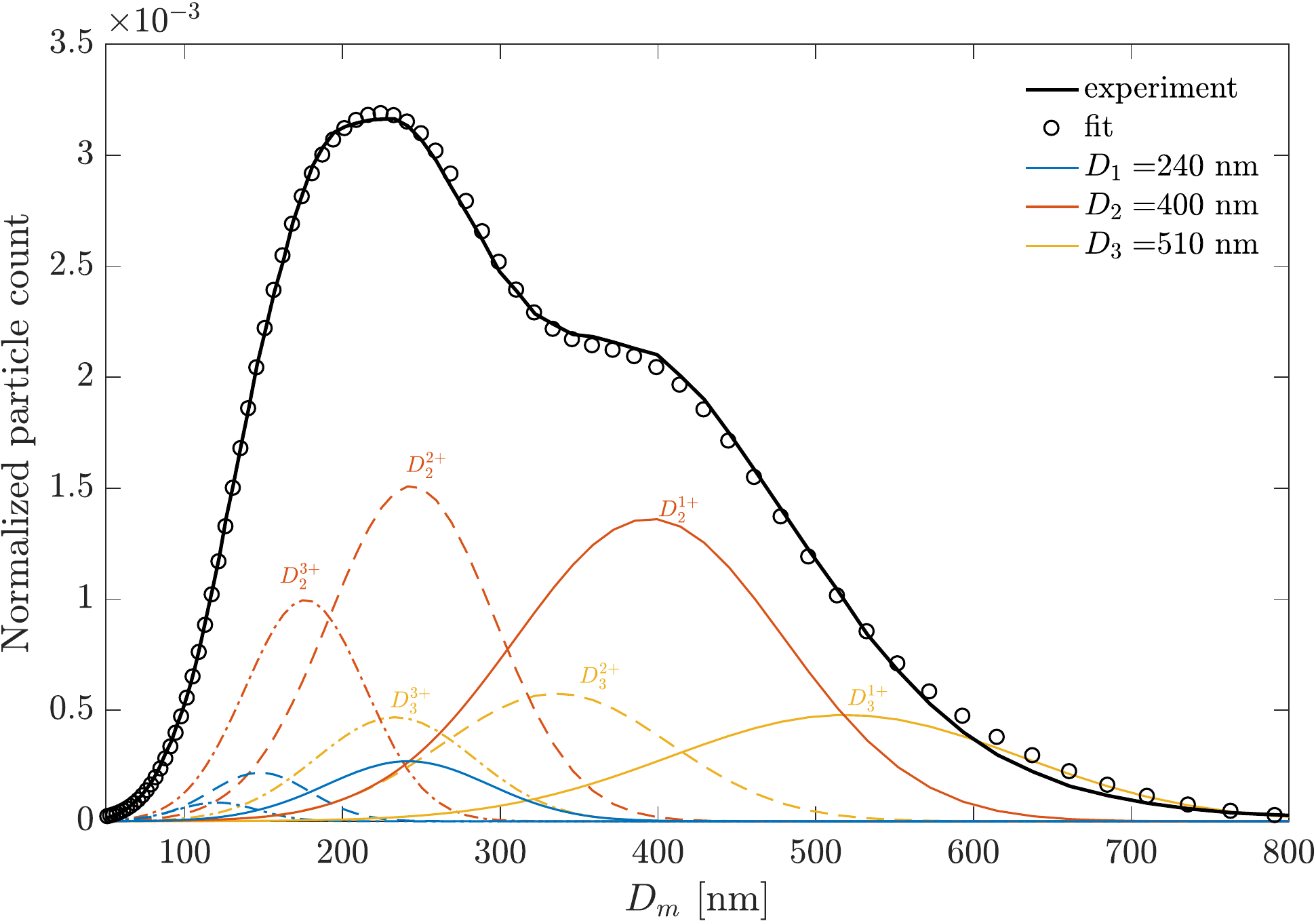}% Here is how to import pdf art
	\caption{\label{fig:fit_uncharged} Plot showing measured (circles) and fitted (solid black line) droplet mobility diameter distribution from which the size distribution is extracted. The distribution is well represented by a fit using three droplet diameters $D_1$ (blue), $D_2$ (red) and $D_3$ (yellow) and charge states of one (solid colored lines), two (dashed colored lines) and three (dashed-dotted colored lines).}
\end{figure}

\paragraph*{Droplet size:}
The droplet size distribution exiting the DMA can be obtained from the SMPS measurement with the Xray neutralizer \#2 turned on. Fig. \ref{fig:fit_uncharged} shows a measured droplet mobility diameter ($D_m$) distribution which is constant regardless of the charger setting as the neutralizer produces an equilibrated distribution of charges. A multi-Gaussian component fit was performed to extract the size and abundance of droplets in the aerosol flow with the i-th size component of the distribution in charge state $q$ centered around $D_m = D_i^{q}$. The distribution of the smallest singly charged droplets present in the sample is thus represented by a Gaussian centered at $D_m = D_1^{1+}$. By employing the DMA transfer and aerosol neutralization theories \cite{Wiedensohler1988,Knutson1975,Stolzenburg1988}, additional components are fitted to account for larger droplets with similar mobility ($D_2^{2+}$ and $D_3^{3+}$ ) and their Boltzmann-equilibrated charge states ($D_1^{2+,3+}$, $D_2^{1+,3+}$, $D_3^{1+,2+}$). A good fit to the experimental distribution was obtained by using three size components with mobillity diameters $D_i^{1+}=$240 nm, 400 nm and 510 nm and accounting for up to 3 (positive) charges after neutralization. The relative abundances of $D_i^{q}$ are used to calculate an average droplet charge $<q>$. 
The peak width of each Gaussian component is assumed to increase linearly with $D_i^{q}$. The full-widths at half maximum obtained for singly-charged droplets are $\sim$120, 200 and 255 nm for $D_1^{1+}=$240 nm, $D_2^{1+}=$400 nm and $D_3^{1+}=$510 nm, respectively. These values are about 3 times higher then the theoretical estimate of the ideal DMA transfer function, indicating the regime in which non-ideal behavior needs to be considered. For example, due to the high aerosol concentration ($>10^7$ particles/cm$^3$) and low sheeth-to-aerosol flow ratio ($<5$) space-charge effects can become significant and lead to mobility shifts and distribution broadening \cite{Alonso1996,Alonso2000,P.Camata2001}.

\begin{figure}[b]
	\centering
	\begin{minipage}{0.49\columnwidth}
	\includegraphics[width=0.9\columnwidth]{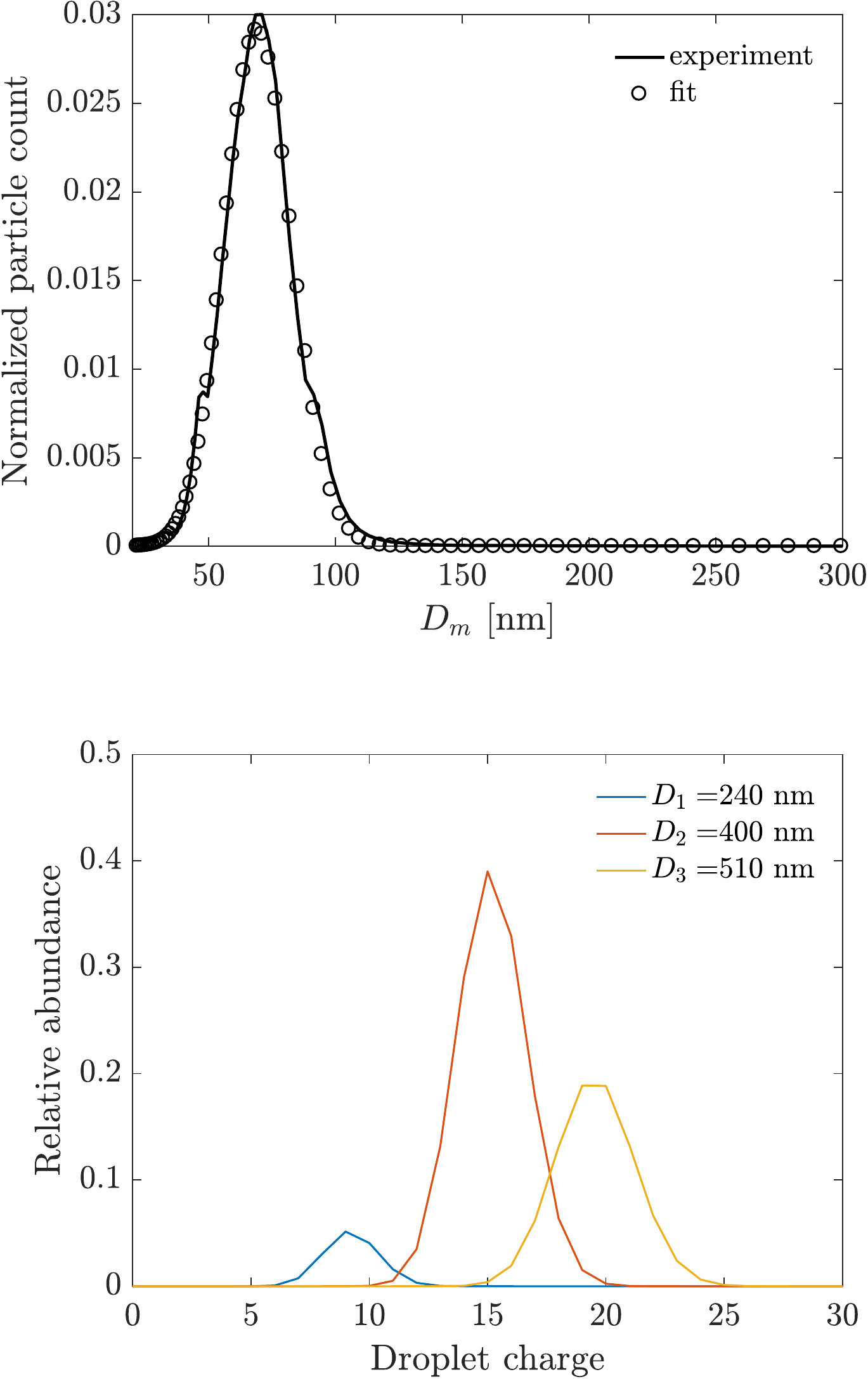}% Here is how to import pdf art
	\caption{\label{fig:fit_charged35+} Mobility (top) and charge distribution (bottom) for charger current set to (+)35 nA. Charging product is $N_it=1.12\times10^{13}$ ions m$^{-3}$s. The weighted average charge is $<q>=16$.}
	\end{minipage}
	\hfill
	\begin{minipage}{0.49\columnwidth}
	\includegraphics[width=0.9\columnwidth]{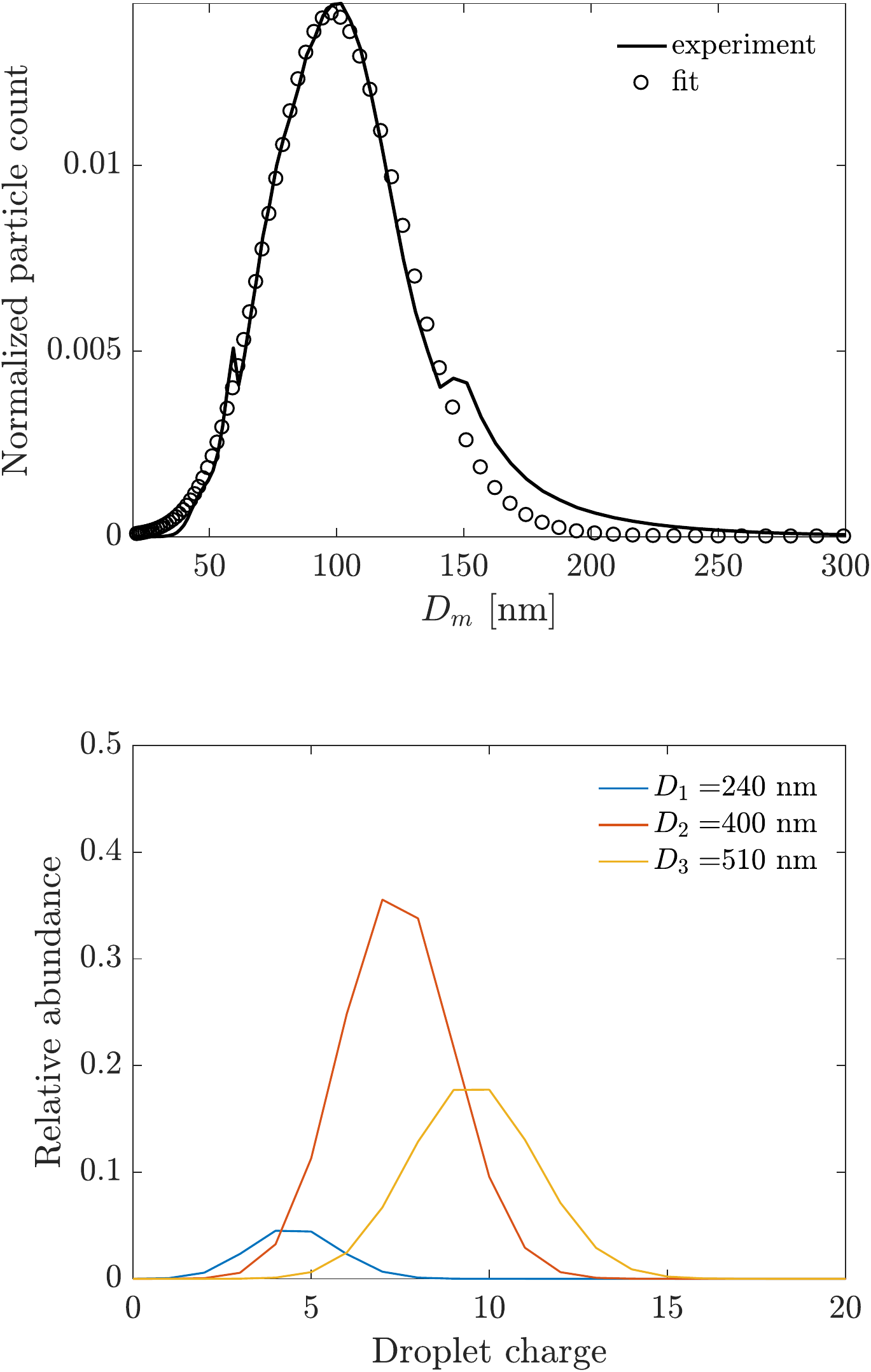}% Here is how to import pdf art
	\caption{\label{fig:fit_charged01+} Mobility (top) and charge distribution (bottom) for charger current set to (+)0.1 nA. Charging product is $N_it=1.80\times10^{12}$ ions m$^{-3}$s. The weighted average charge is $<q>=8$.}
	\end{minipage}
\end{figure}

\begin{figure}
	\centering
	\begin{minipage}{0.49\columnwidth}
	\includegraphics[width=0.9\columnwidth]{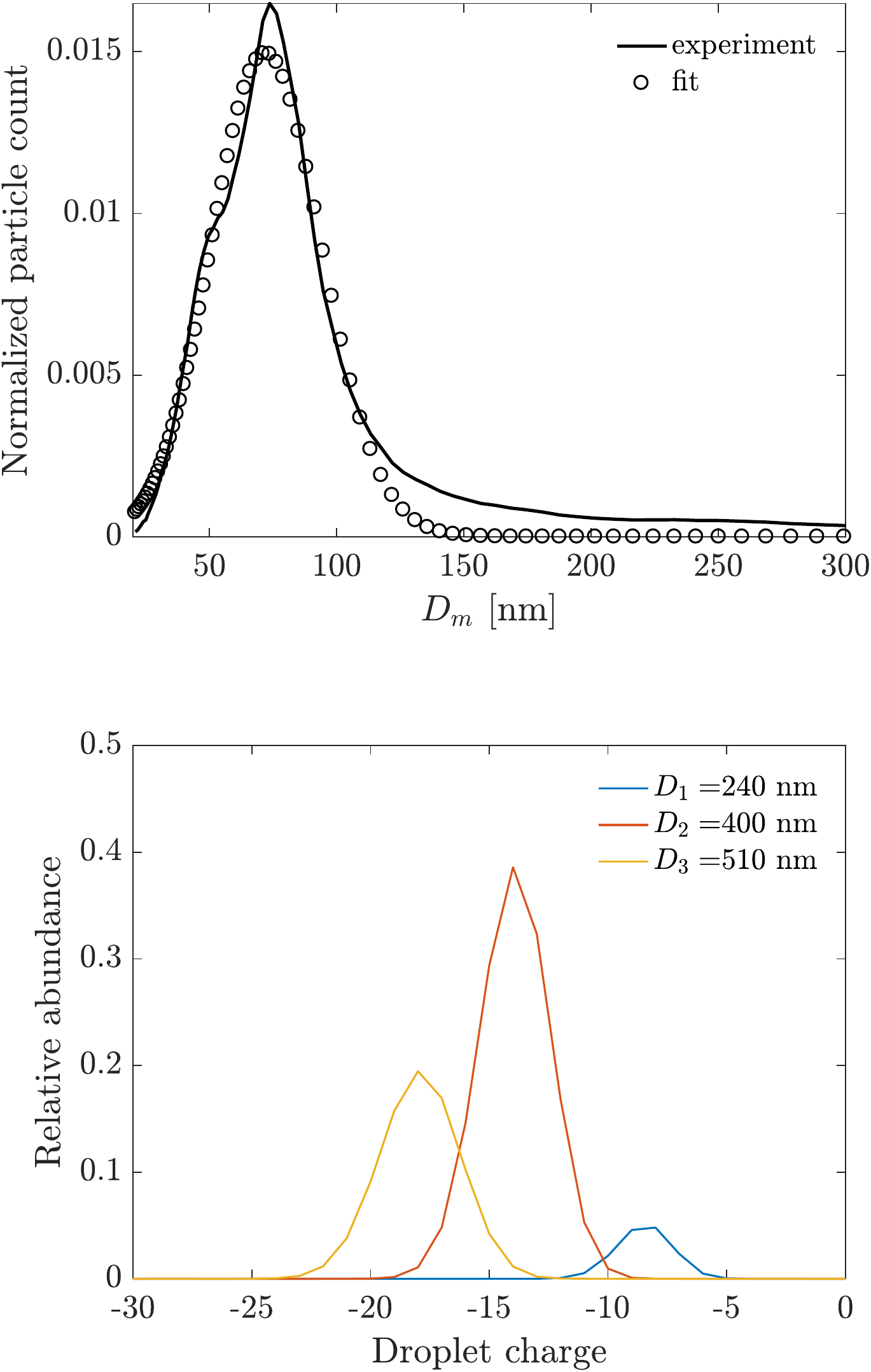}% Here is how to import pdf art
	\caption{\label{fig:fit_charged35-} Mobility (top) and charge distribution (bottom) for charger current set to (-)35 nA. Charging product is $N_it=8.35\times10^{12}$ ions m$^{-3}$s. The weighted average charge is $<q>=-15$.}
	\end{minipage}
	\hfill
	\begin{minipage}{0.49\columnwidth}
	\includegraphics[width=0.9\columnwidth]{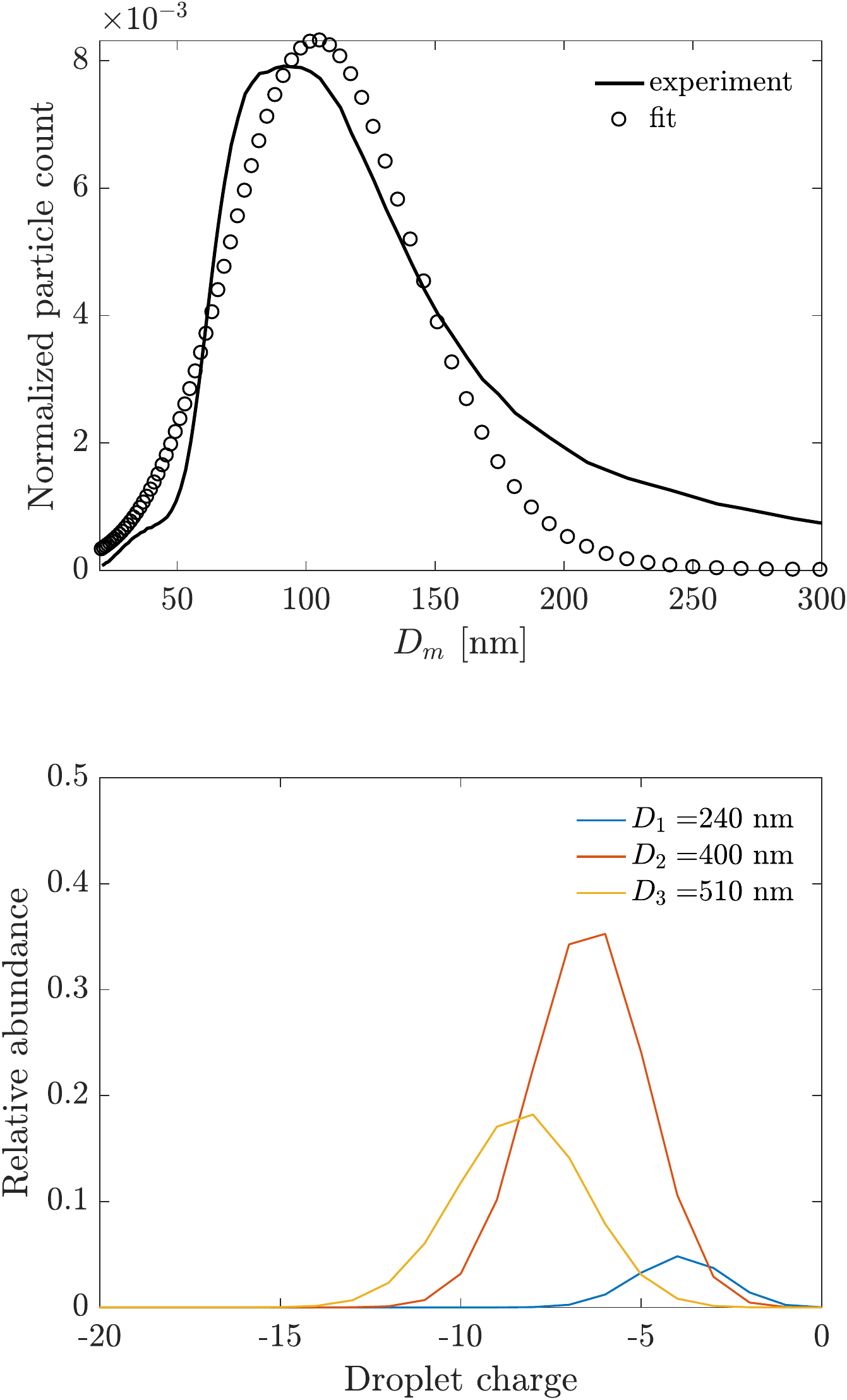}% Here is how to import pdf art
	\caption{\label{fig:fit_charged2-} Mobility (top) and charge distribution (bottom) for charger current set to (-)2 nA. Charging product is $N_it=1.41\times10^{12}$ ions m$^{-3}$s. The weighted average charge is $<q>=-7$.}
	\end{minipage}
\end{figure}

\clearpage
\paragraph*{Droplet charge:}
When the aerosol neutralizer \#2 is turned off, the mobility distribution measured by the SMPS contains information on the droplet charge. A multi-Gaussian component fit can be used to extract the charge distribution as a function of the droplet size. This fit is based on the unipolar-charging theory which has been extensively investigated in the literature \cite{Zheng2017}. Here, a stochastic approach to the diffusion-charging mechanism is used to obtain not only the average charge, but the complete charge distributions for each droplet diameter \cite{Boisdron1970,Biskos2004,Biskos2005}. It is based on solving the differential-difference equations describing the charging process by using only two quantities; charge flux onto a particle ($J$) and the product $N_it$, where $N_i$ is the ion density and $t$ is the residence time in the charger. A comprehensive review of this method can be found in Refs. \cite{Biskos2004,Biskos2005}. 
The charge flux $J$ depends on the exact charging conditions and many theories have provided various mechanisms for determining $J$. Considering the droplet size range and the carrier gas (N$_2$) used in this work, continuum charging theory developed independently by a number of authors \cite{Arendt1926,Pauthenier1932,Fuchs1947,Bricard1949,Gunn1954} seems to be the best choice. The charge flux onto a particle with diameter $D_i$ and charge $n$ is then given (in SI units) as
\begin{equation}
J = K_E\frac{4\pi \mathbf{D}_\mathrm{ion}N_ine^2}{kT\left[\exp(K_E\frac{ne^2}{D_ikT})-1\right]}, \label{eq:continuum_flux} 
\end{equation} 
where $K_E=1/4\pi\varepsilon_0$, $\mathbf{D}_\mathrm{ion}$ is the ion diffusion coefficient, $e$ the elementary charge, $k$ the Boltzmann constant and $T$ is the temperature. This equation considers only the Coulomb force, while the image force can normally be neglected for submicron-sized droplets in the continuum regime \cite{Zheng2017}. Therefore, knowing the $N_it$ product would allow us to determine the droplet charge distribution for each droplet size present in the sample. The charging parameter $N_it$ is obtained from the fit to the measured mobility distribution, employing the droplet sizes ($D_1$, $D_2$ and $D_3$) and relative abundance information determined from the size-determination routine. 
Fit results for the relevant charger settings are shown in Fig. \ref{fig:fit_charged35+}-\ref{fig:fit_charged2-}. For the positive case, the fit to the mobility distributions is in very good agreement with the measurement and shows that VMI images recorded correspond to an average droplet charge of $<q>=$ 8 and 16. The average charge is obtained from a weighted average of the charge distribution employing relative abundances determined above. The shoulders visible at $D_m\sim50$ and 150 nm are a signature of the high droplet concentration which leads to saturation and coincidence counting of particles in the SMPS.
In the negative case, the average charge states obtained from the fit are $<q>=$ -7 and -15. In the case of the lowest negative charging setting the fit agrees less well with the measured mobility distributions showing a tail extending towards larger $D_m$ values. Since two distinct chargers were used for each polarity, it is likely that the negative charger does not perform reliably for the lowest charging current (Fig. \ref{fig:fit_charged2-}). 

As for the neutral droplet sample, it is important to note that it is neutral only on average. The current experimental setup does not allow for precise characterization of the charge distribution of the neutral sample. This could in principle be achieved by using a variable-polarity SMPS. Our simulations (Fig. 2b in the main text) for the eKE spectrum in the region below 0.1 eV, which is very sensitive to the exact charge state, indicate that a small amount of positively charged particles might be present in the sample. This is not unexpected given the non-optimal conditions in the DMA (see above). 

\clearpage
\section{ Electron scattering model \label{sec:SImodel} }
\subsection{Genuine electron binding energy spectrum}
\begin{figure}[h]
	\includegraphics[width=0.6\columnwidth]{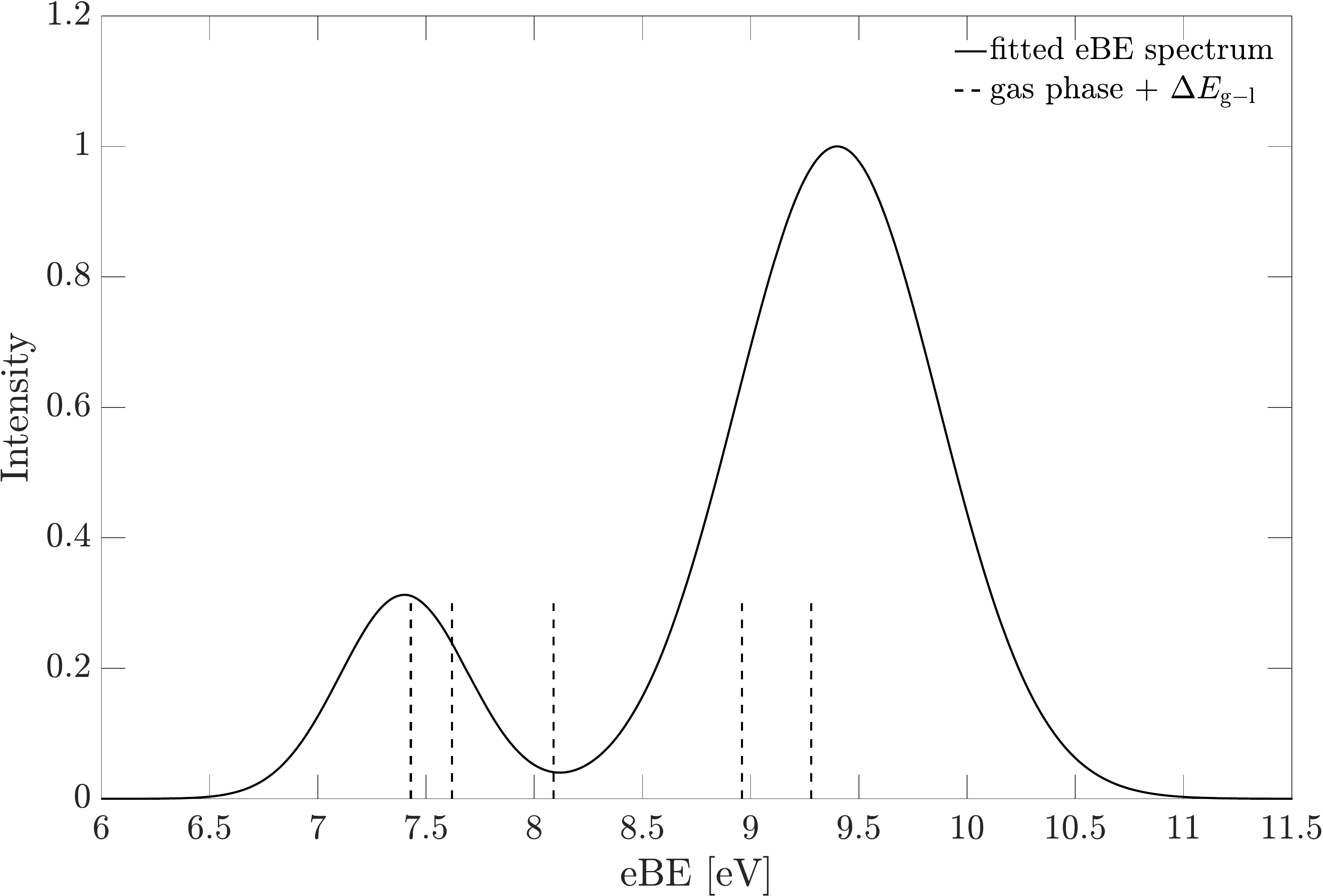}% Here is how to import pdf art
	\caption{\label{fig:ebe_genuine_gas} Genuine eBE spectrum of liquid DEHP obtained from a fit of two Gaussians to the experimental eKE spectrum of neutral droplets. The eBE spectrum is compared with ionization energies calculated at the B3LYP/6-311++g** level of density functional theory for gas phase DMP shifted by 2 eV towards lower energy.}
\end{figure}

The result of the first step of the photoionization, the excitation of an electron into the conduction band of the droplet, is described by the genuine properties of the system, i.e. its genuine binding energy (eBE) spectrum and its genuine PAD. We assume the latter to be isotropic, which leaves the condensed phase genuine eBE spectrum of DEHP to be determined. Photoelectron spectra at high photon energies can sometimes yield a reasonable  approximation to the genuine eBE spectrum. Such experiments were performed for liquid and solid benzene \cite{Schwarz1982,Scott1982,Saik1994,Takahashi1981}, but no data is available for DEHP. Therefore, the genuine eBE spectrum of liquid DEHP is determined by fitting it to the experimental eKE spectrum of neutral droplets (Fig. 2b in the main text) in a two-step procedure. We only used experimental data for eKE$>$0.1 eV. Below 0.1 eV the eKE spectrum is too sensitive to the exact charge state, which is not known exactly as the neutral droplet sample is neutral only on average. The simulations (Fig. 2b in the main text) hint that a small amount of positively charged particles might be present in the sample as a consquence of the non-optimal conditions in the DMA (see above). For the purposes of fitting the genuine eBE spectrum the interface potential parameters (barrier height and width, see below) are fixed at $V_0=1$ eV and $w=0.1$ nm. 
In the first step, the high eKE region is fitted assuming a single Gaussian band for the genuine eBE spectrum (Fig. \ref{fig:ebe_genuine_gas}). Band position and width are well constrained to within better than 0.1 eV by the signal onset in the experimental eKE spectrum. In the following step, the complete eKE spectrum above 0.1 eV is fitted allowing for a second Gaussian band at higher eBE (Fig. \ref{fig:ebe_genuine_gas}). The resulting bands are only to be considered as an effective representation of the true genuine eBE spectrum, which probably contains contributions from several electronic levels. They become difficult to distinguish as band broadening upon condensation reduces the number of distinguishable bands. For example, in solid benzene the number of resolved bands reduces from 8 (gas) to 4 (solid) \cite{Takahashi1981}. The band positions we determined are in reasonable agreement with a theoretical photoelectron stick-spectrum (showing vertical binding energies) of gas phase dimethyl phthalate (DMP) calculated using density functional theory and applying a condensation shift of $\Delta E_\mathrm{g-l}\sim 2$ eV (Fig. \ref{fig:ebe_genuine_gas}), similar to liquid benzene \cite{Schwarz1982,Scott1982,Saik1994}. Vertical binding energies were calculated for the \textit{trans}-isomer of DMP in its equilibrium geometry using the B3LYP functional with the 6-311++G** basis set as implemented in the Gaussian software package \cite{Frisch2014}. Excited states of the ion were obtained by adding the corresponding difference of Kohn-Sham orbital energies of the neutral.

\clearpage
\subsection{DEHP scattering cross sections}

\begin{figure}[h]
	\includegraphics[width=0.6\columnwidth]{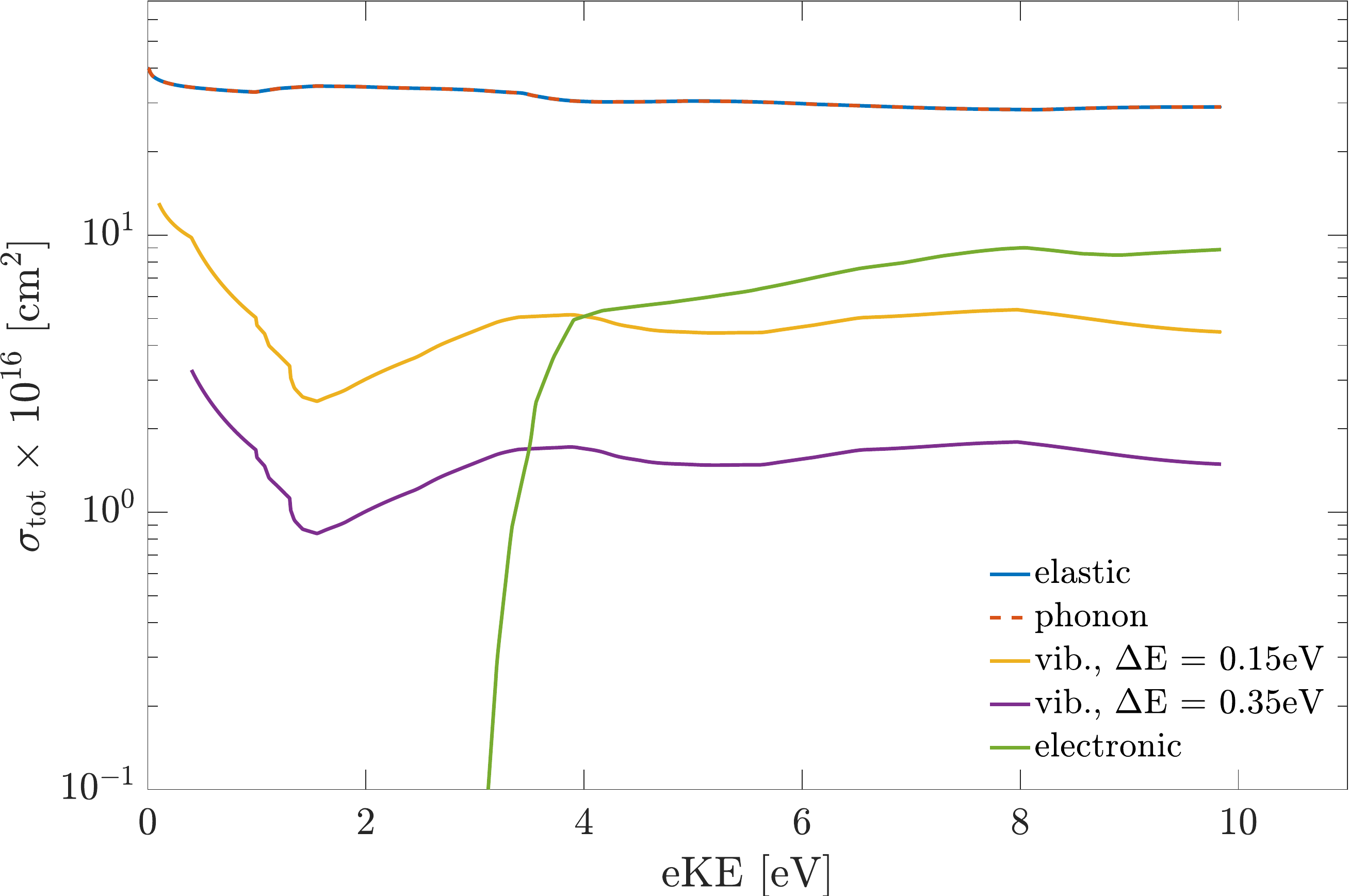}% Here is how to import pdf art
	\caption{\label{fig:cross_sections} Differential scattering cross section used to describe the electron transport scattering in DEHP. Elastic and phonon channels have equal cross section with $\Delta E=$0 and 25 meV, respectively. Here, eKE refers to the kinetic energy of the electron inside the droplet.}
\end{figure}

Describing the second step in the photoionization (transport scattering) requires the knowledge of scattering cross sections. Detailed information, however, about energy loss and angular characteristics of different molecular scattering channels, such as the differential scattering cross sections available for amorphous ice \cite{Michaud2003} and liquid water \cite{Signorell2016,Luckhaus2017g}, are scarce in the literature. Hydrocarbon thin-films have been studied in the past so that a reasonable amount of data on electron scattering in hydrocarbon films is available \cite{Sanche1982,Sanche1979b}. Of the substances for which data is available, solid benzene comes closest to DEHP. The work on benzene \cite{Goulet1985,Goulet1986} provides differential scattering cross sections (only energy loss, no angular information) for the electronic ($\sigma_\mathrm{elec}$), vibrational ($\sigma_\mathrm{vib}$) and quasi-elastic ($\sigma_\mathrm{el}$ and $\sigma_\mathrm{phon}$, phonon and elastic) channels in the energy range from 1 to 10 eV. We used these data as a basis to construct the following model for electron scattering in DEHP droplets (Fig. \ref{fig:cross_sections}). 
In the energy range probed in this work ($<4$ eV), electronic scattering channels only contribute to the scattering at the highest kinetic energies, effectively corresponding to electron loss ($\Delta E_\mathrm{elec}=3.9$ eV). This hardly affects the appearance of the observed eKE spectrum.
The vibrational contribution $\sigma_\mathrm{vib}$ is split into two channels ($\Delta E_\mathrm{vib}=$0.1 and 0.4 eV) to account for the two broad vibrational bands of solid benzene. The energy losses of the vibrational channels are adjusted to $\Delta E_\mathrm{vib}=$0.15 and 0.35 eV to match the IR spectrum of liquid DEHP \cite{NIST}. $\sigma_\mathrm{el}$ and $\sigma_\mathrm{phon}$, which could not be resolved in the experiments on benzene, were obtained by splitting the contribution of the single quasi-elastic channel in benzene equally between the elastic channel ($\sigma_\mathrm{el}$, $\Delta E_\mathrm{el}=$0) and the phonon scattering channel ($\sigma_\mathrm{phon}$). For the latter we estimated an energy loss of $\Delta E_\mathrm{phon}=$25 meV based on the typical energy-loss spectra of benzene and long-chain hydrocarbons \cite{Sanche1982,Sanche1979b,Goulet1985,Goulet1986}. 
While $\sigma_\mathrm{vib}$ is not expected to differ greatly between DEHP and benzene, $\sigma_\mathrm{phon}$ might be modified by the two additional 10-carbon chains of DEHP. However, sensitivity tests show no significant effect on eKE spectra when varying the ratio of $\sigma_\mathrm{el}$ to $\sigma_\mathrm{phon}$ between 0.3 and 3. Similarly, varying the energy loss of $\sigma_\mathrm{phon}$ between $\Delta E_\mathrm{phon}=$10 and 50 meV has no significant influence on the eKE spectra. Finally, the cross sections for eKE $<1$ eV (only relevant when $V_0<$1 eV) are obtained by double-logrithmic extrapolation. 

\clearpage
\subsection{Scattering, Electron escape and Propagation}
We employ a probabilistic description of transport scattering in terms of a random sequence of localized scattering events, which amounts to a Monte-Carlo solution of the transport equation, followed by transmission through the droplet-vacuum interface (escape) and detection (VMI).
The initial distribution of electrons in the conduction band is given by the genuine eBE spectrum and genuine PAD (here assuming an isotropic velocity distribution) with the spatial distribution across the droplet determined by the local laser field intensity. The latter is calculated by solving Maxwell's equations numerically within the Distributed Dipole Approximation \cite{Yurkin2011}. The probability of forming an electron in the conduction band by 2-photon excitation is proportional to the square of the local light intensity.
The trajectory of the electron through the droplet follows a random walk between scattering events with the distribution of step lenghts, energy losses and deflection angles given by the energy dependent differential cross sections \cite{Signorell2016,Luckhaus2017g}.

For an ideal homogeneous spherical droplet with an infinitely narrow barrier and a  perfectly smooth surface, angular momentum conservation would lead to Snell's law of refraction for the (classical) transmission of the electron, with total reflection, depending on the angle of incidence $\theta$ w.r.t. the surface normal. We had found previously that a more realistic description of the (classical) escape from a small droplet replaces the total reflection of an electron by an inelastic scattering event at the surface, with an energy loss equal to the barrier height and a preferential deflection in the forward direction \cite{Signorell2016}. The rationale was that electrons hitting the interface at large $\theta$ are more prone to experience the molecular scale inhomogeneity and roughness of the droplet surface. 

For the present study we have extended our previous model to account for quantum effects (tunnelling, above-barrier reflection) and the influence of surface charge on the electron's escape from the particle (step 3 in Fig. 1 in the main text). 
Both are governed by the effective potential function
\begin{eqnarray}
V(r) &=& \frac{V_0}{2} \left\{1 + \tanh \left[\frac{2a}{w} \left(r-R_D+\frac{w}{2}\right) \right] \right\} - \frac{q \cdot e^2}{4\pi\varepsilon_0 R}  + \frac{L^2}{2m_eR^2},\label{eq:potential}
\end{eqnarray}
where $\varepsilon_0$ is the vacuum permittivity, $e$ the elementary charge, and $m_e$ the electron's mass. $q$ is the droplets charge state, $R_D$ its radius, $r$ the radial distance from its center, and $R=\mathrm{max}(r,R_D)$. 
The first term describes the bare potential barrier with height $V_0$ and width $w$. For the neutral particle $V_0$ specifies the position of the vacuum relative to the bottom of the conduction band. The barrier width $w$ is defined w.r.t. the potential as the width over which the barrier reaches 99.9\% of $V_0$ and therefore the parameter $a=\tanh^{-1}(0.999)$.   
The second term is the potential arising from the droplet's charge  $q\cdot e$, which we assume to be uniformly distributed on the surface \cite{Ziemann1995}. Inside the particle this term is constant, while outside it takes the form of a Coulomb potential shifting the vacuum level relative to the bottom of the conduction band by $q \cdot e^2 / 4\pi\varepsilon_0 R_D$.
The last term is the centrifugal potential arising from the electron's angular momentum $L$. We neglect the variation of the centrifugal potential inside the droplet since the de Broglie wavelength of the escaping electron is negligible compared with $R_D$. $L$ is determined by the electron's kinetic energy $E_k$ and angle $\theta$ relative to the surface normal after the inelastic forward scattering event at the surface

\begin{eqnarray}
L^2 &=& 2m_eE_k R_D^2 sin^2\theta.
\end{eqnarray}

The probability for the electron to escape once it has reached the droplet's surface - the transmission probability - is calculated from the numerical  solution of the radial Schr{\"o}dinger equation for the effective potential given by Eq.\ref{eq:potential}. The main difference to the previous classical treatment is the finite probability for an electron to be reflected back into the droplet even if its kinetic energy exceeds the escape barrier (above-barrier reflection, see below). If the droplet is uncharged the electron's velocity vector following transmission is directly projected onto the detector. The Coulomb force exerted by charged droplets, however, leads to an additional deflection of the electron after escape. This is treated classically to yield the final direction of motion of the electron given by \cite{Landau}
\begin{eqnarray}
	\theta_f &=& \int_{R_D}^{\infty} \frac{L/r^2}{\sqrt{2m_e\left[E_k-V(r)\right]}}\mathrm{d}r. \label{eq:deltaT}
\end{eqnarray}

Here $\theta_f$ is the angle between the electron's final velocity vector and the surface normal at its point of escape and $V(r)$ is given by Eq.(\ref{eq:potential}).

Typically, $10^8$ to $10^9$ Monte Carlo trajectories are averaged for a given droplet size and charge state. For the direct comparison with the experiment, the results for the three droplet diameters ($D_1$, $D_2$ and $D_3$) are averaged over different charge states according to the experimentally determined weights of the components $D_i^q$ of the size and charge distributions. 

\clearpage
\section{Results}
\subsection{Determination of the high kinetic energy onset \label{sec:SIonset} }
\begin{figure}[h]
	\includegraphics[width=0.6\columnwidth]{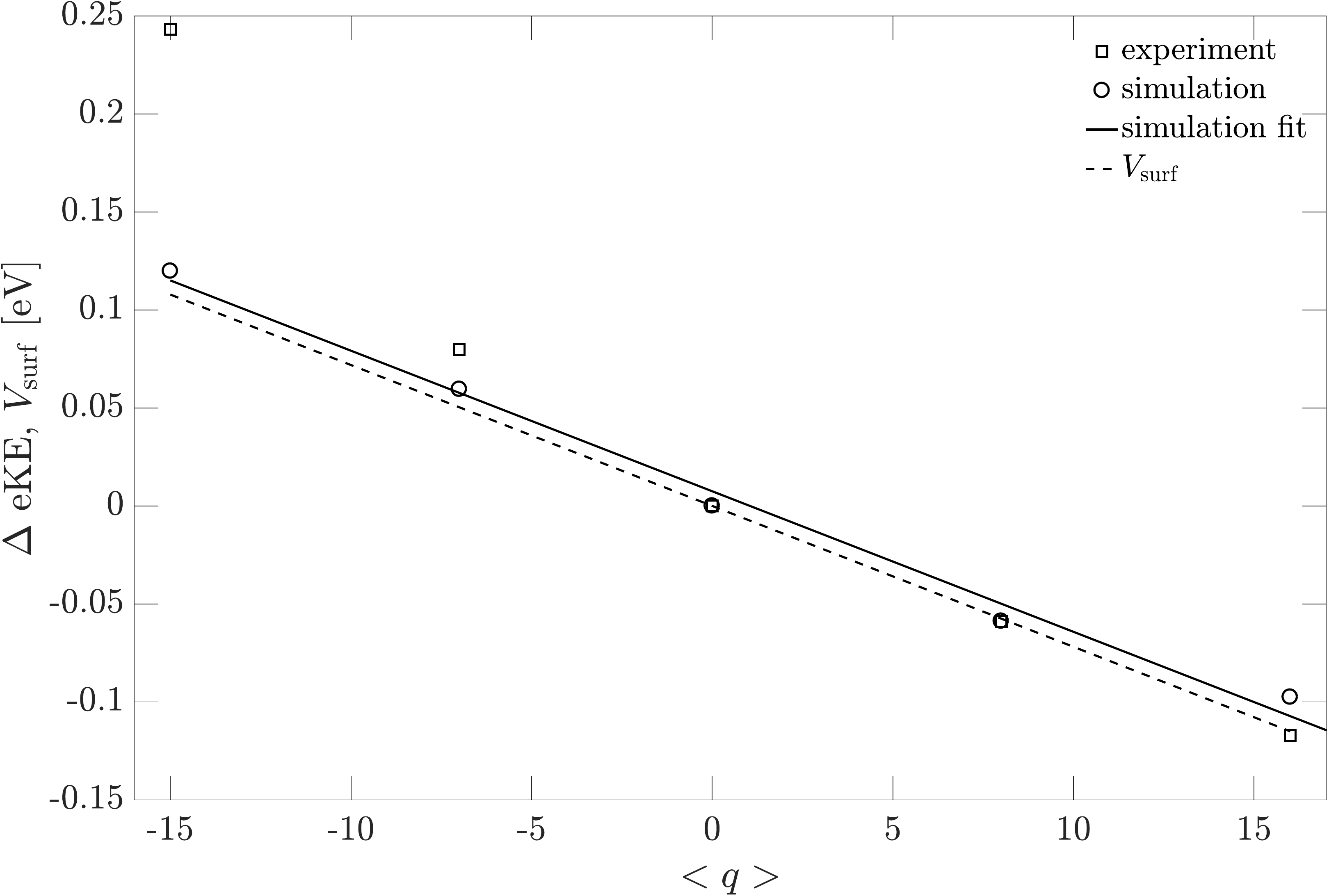}% Here is how to import pdf art
	\caption{\label{fig:ebe_onset_fit} $\Delta$eKE as a function of the average droplet charge $<q>$ determined from the experiment (squares) and simulation (circles). Linear fit to the simulated data (solid line) is in good agreement with the average surface charge potential term (dashed line) $-<q>\cdot e^2 / 4\pi \varepsilon_0 R_D$ for $R_D$=200nm.}
\end{figure}

The determination of the eKE onsets (the highest eKE observed in the spectrum) is in general not straightforward in condensed phase experiments \cite{Roy2018}. Here, we use a simple definition merely to illustrate the effect of droplet charge on the spectrum. The onsets were determined as the 1/$e^2$ signal of the second band at eKE$\sim 1.5$ eV (squares). Fig. \ref{fig:ebe_onset_fit} shows experimental (squares) and simulated (circles) eKE shifts relative to the uncharged case ($\Delta$eKE) as a function of charge $<q>$.  A linear fit (solid line) to the simulations (circles) results in a slope of $-7.2$ meV, which agrees with the slope predicted from the surface charge potential term in Eq.\ref{eq:potential} for a droplet diameter of 200 nm (dashed line), indicating the consistency of size and charge distributions. Except for  $<q>=$ -15, the experiment (squares) closely follows the linear dependence on $<q>$.

\clearpage
\subsection{Photoelectron angular distributions \label{sec:SIangular} }
\begin{figure}[h]
	\includegraphics[width=0.8\columnwidth]{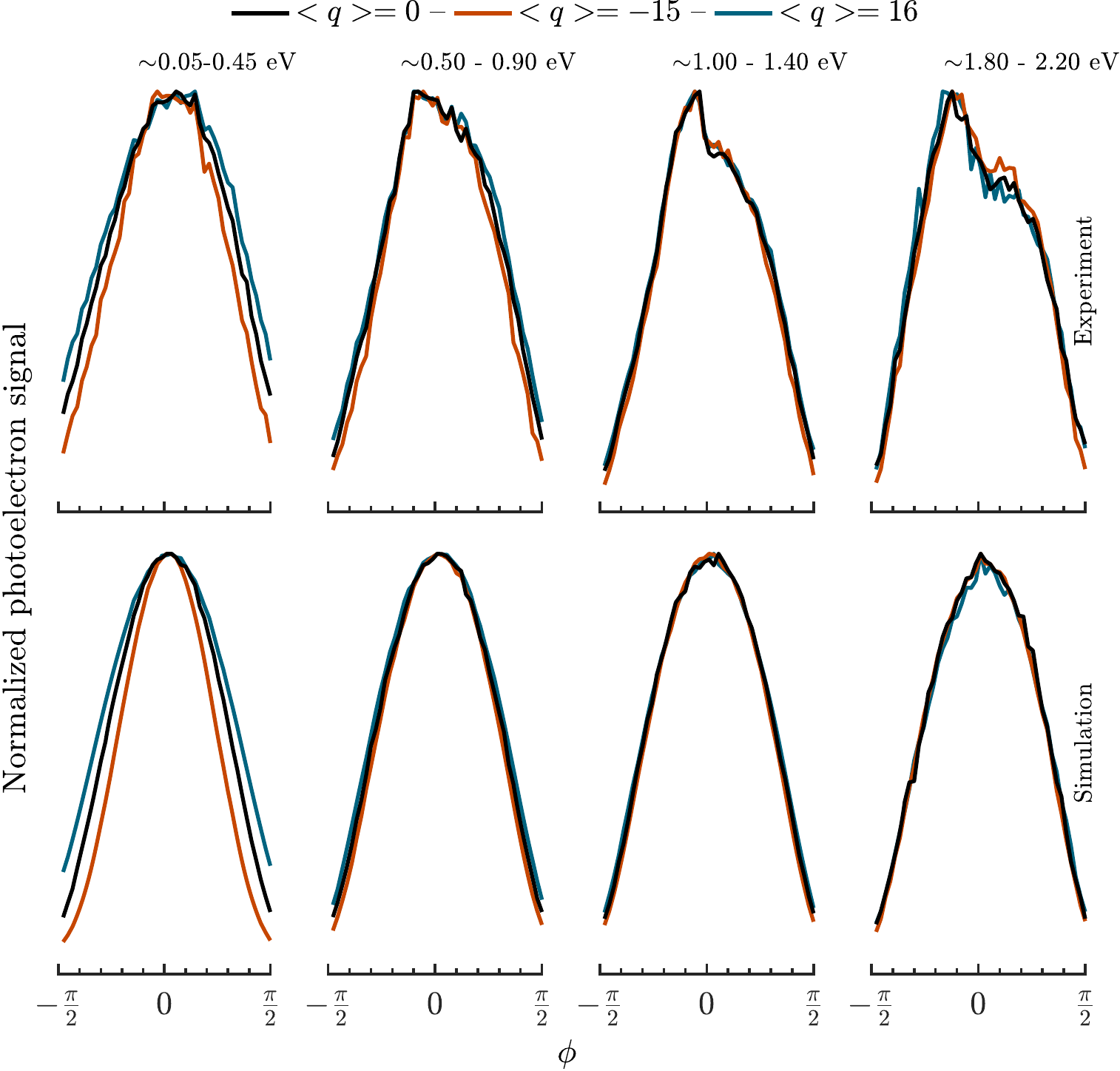}% Here is how to import pdf art
	\caption{\label{fig:angular} Integrated angular distributions of the photoelectron signal corresponding to different eKE ranges as a function of $\phi$  for neutral (black line), negatively (red line) and positively (blue line) charged droplets for $V_0=1$ eV and $w=0.1$ nm. The approximate eKE ranges for the experimental (top panels) and simulated (bottom panels) distributions are indicated in the top panel.}
\end{figure}

The PADs are analyzed in terms of angular distributions obtained by integration of the experimental and simulated images. Fig. \ref{fig:angular} shows the experimental (top panels) and simulated (bottom panels) photoelectron signal as a function of the angle $\phi$ for four different energy regions. $\phi$ is defined with respect to the light propagation direction, so that $\phi=0$ and $\phi=\pi/2$ correspond to laser propagation and polarization directions, respectively. The integration extends over a given range of the electrons velocity in the detector plane specified in terms of the corresponding eKE values. The maximum in the angular distributions for higher eKEs arises from inhomogeneities of the electron detector and/or aberrations in the VMI optics.

\clearpage
\subsection{Electron transmission probability}
\begin{figure}[b]
	\centering
	\begin{minipage}{0.45\columnwidth}
		\includegraphics[width=\columnwidth]{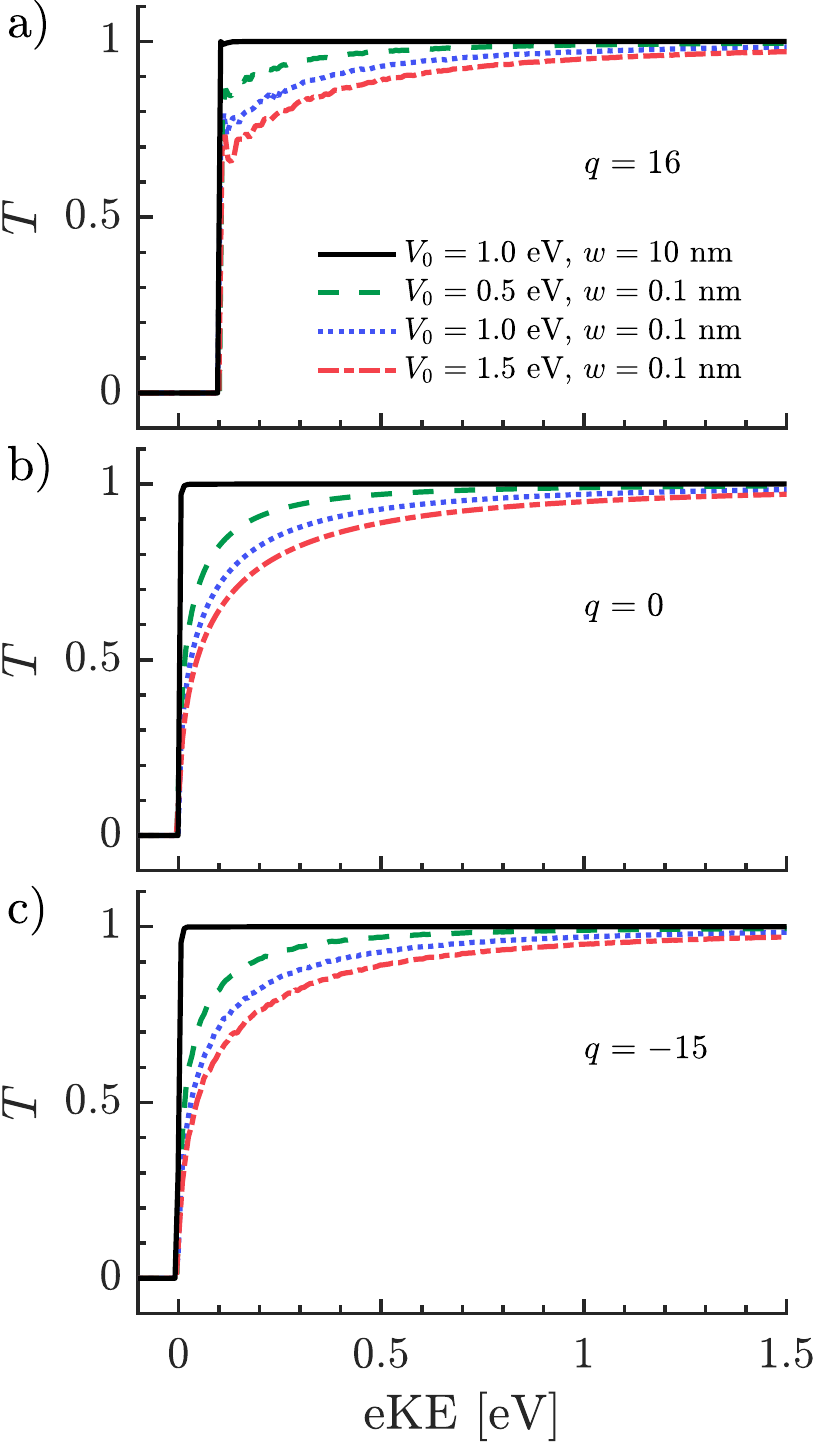}% Here is how to import pdf art
		\caption{\label{fig:transmission} $T$ as a function of the eKE (inside the droplet) for electrons with $\theta=0^\circ$ and different interface potential parameters ( $w$ and $V_0$) for droplet charge states with $q=$+16 (a), 0 (b) and -15 (c).}
	\end{minipage}
	\hfill
	\begin{minipage}{0.45\columnwidth}
		\includegraphics[width=\columnwidth]{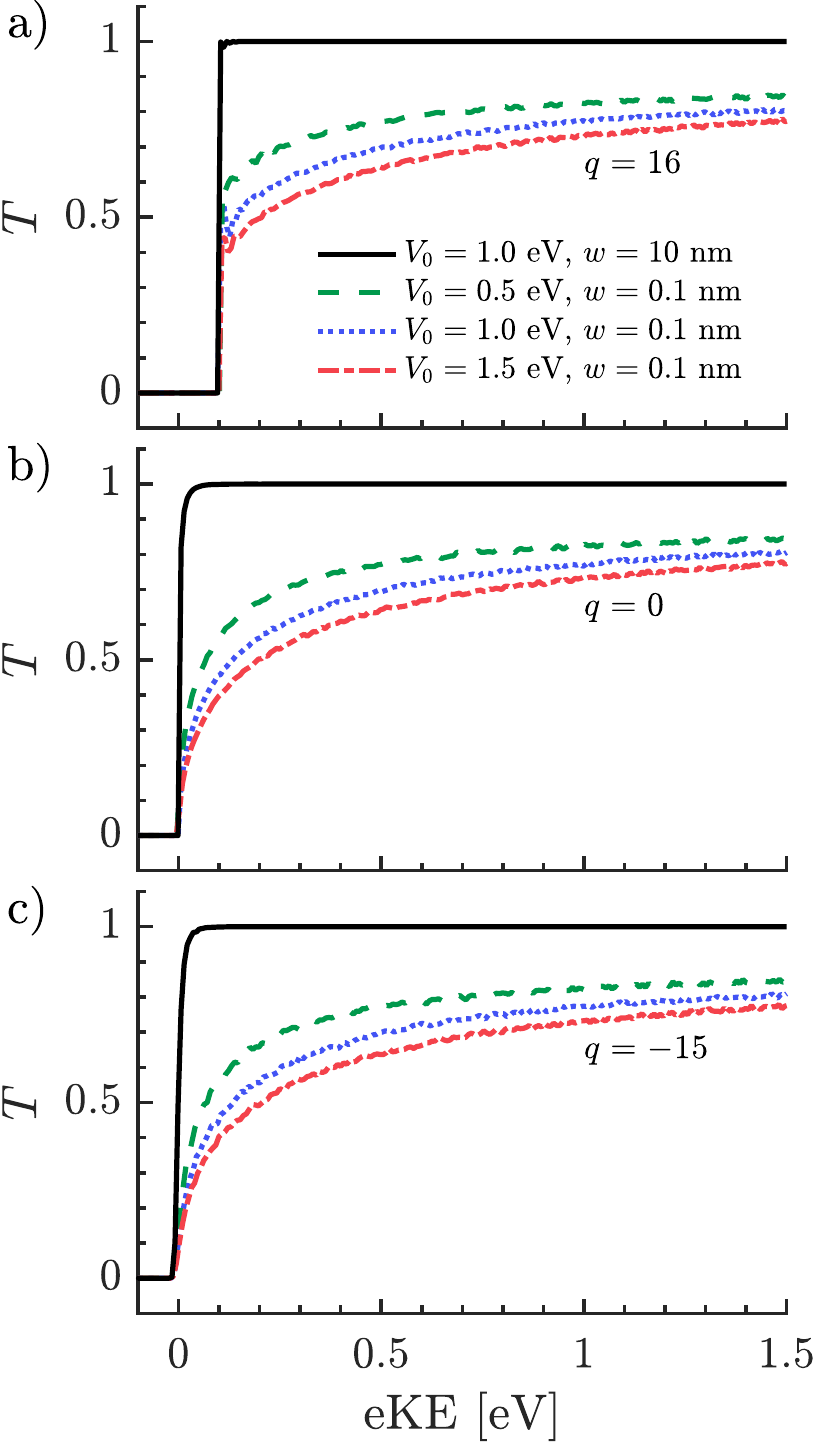}% Here is how to import pdf art
		\caption{\label{fig:transmission1} $T$ as a function of the eKE (inside the droplet) for electrons with $\theta=60^\circ$ and different interface potential parameters ( $w$ and $V_0$) for droplet charge states with $q=$+16 (a), 0 (b) and -15 (c).}
	\end{minipage}
\end{figure}

The electron transmission probability is calculated from the solution of the radial Schr{\"o}dinger equation for the effective potential $V(r)$ given by Eq.\ref{eq:potential}. For the numerical solution we follow the approach of reference \cite{Ando1987}. Fig. \ref{fig:transmission} and \ref{fig:transmission1} show the charge-dependent (panels a, b and c) transmission coefficient $T$ as a function of the photoelectron kinetic energy (eKE) for electron incidence angles $\theta=0^\circ$ (normal incidence) and 60$^\circ$, respectively. $\theta$ is the angle between the direction of motion of the electron and the surface normal at the point of escape. 
For a barrier width of $w=10$ nm, above-barrier reflection becomes insignificant. In this classical limit, $T$ as a function of eKE is largely independent of the barrier height $V_0$. Therefore, only the values for $V_0=1$ eV are shown (black solid line). For a narrow barrier of $w=0.1$ nm, however, the above-barrier reflection is pronounced. Regardless of the average charge state $<q>$, the highest $T$ is obtained for the lowest $V_0$. This is expected since the kinetic energy range where above-barrier reflection is significant scales with the barrier height. By the same argument $T$ decreases with increasing $\theta$. For electrons impinging on the surface with larger $\theta$, the growing centrifugal potential term in Eq.\ref{eq:potential} reduces the kinetic energy in the radial coordinate for a given eKE thus increasing the eKE range of significant above-barrier reflection.
The effect of droplet charge itself results in a gradual reduction of $T$ when going from positive to negative charging of the droplet (taking the increase of the effective barrier height by a positive surface charge potential into account). Similarly to a broader barrier, the smooth Coulomb tail that positive charges add to a step barrier tend to quench above-barrier reflections making the transmission behavior more classical. Adding negative charges to a neutral droplet does not change the effective escape barrier, so that the transmission probabilities in the two cases are very similar. Some enhancement of above-barrier reflections might be expected from the increasingly sharp cusp forming at the top of the barrier upon negative charging, but the effect remains insignificant for the small surface charge potentials considered in this study.

\clearpage
\bibliographystyle{apsrev4-1} % Tell bibtex which bibliography style to use
\bibliography{bib_1}% Produces the bibliography via BibTeX.